\documentclass[journal,hidelinks]{IEEEtran}

\usepackage[OT1]{fontenc}
\usepackage[cmex10]{amsmath}
\usepackage{amssymb}
\usepackage{bm}
\usepackage{graphicx}
\usepackage{subfigure}
\usepackage{tabularx}
\usepackage{arydshln}
\usepackage{mathtools}
\usepackage{multirow}
\usepackage{algorithm,algorithmic}
\usepackage{url}
\usepackage{color}
\usepackage{comment}
\usepackage{amsthm}
\usepackage{braket}
\usepackage{physics}
\usepackage{lipsum}
\usepackage{cite}
\usepackage{multicol}
\usepackage{float}
\usepackage{hyperref}
\usepackage[absolute,overlay]{textpos}

\def\ie{\textit{i.e.}~}
\def\eg{\textit{e.g.}~}
\def\etal{\textit{et~al.}~}

\newtheorem{thm}{Theorem}

\makeatletter
\def\ps@IEEEtitlepagestyle{%
  \def\@oddfoot{\mycopyrightnotice}%
  \def\@oddhead{\hbox{}\@IEEEheaderstyle\leftmark\hfil\thepage}\relax
  \def\@evenhead{\@IEEEheaderstyle\thepage\hfil\leftmark\hbox{}}\relax
  \def\@evenfoot{}%
}
\def\mycopyrightnotice{%
  \begin{minipage}{\textwidth}
  \centering \scriptsize
\textcopyright 2024 IEEE. Personal use of this material is permitted.
  Permission from IEEE must be obtained for all other uses, in any current or future
  media, including reprinting/republishing this material for advertising or promotional
  purposes, creating new collective works, for resale or redistribution to servers or
  lists, or reuse of any copyrighted component of this work in other works.
  DOI: \href{https://doi.org/10.1109/TQE.2024.3450852}{10.1109/TQE.2024.3450852}
  \end{minipage}
}
\makeatother
\begin{document}

\title{Quantum Speedup of the Dispersion and\\ Codebook Design Problems}

\author{Kein~Yukiyoshi, Taku~Mikuriya, Hyeon~Seok~Rou,~\IEEEmembership{Graduate Student~Member,~IEEE}, Giuseppe~Thadeu~Freitas~de~Abreu,~\IEEEmembership{Senior~Member,~IEEE}, and Naoki~Ishikawa,~\IEEEmembership{Senior~Member,~IEEE}.
\thanks{K.~Yukiyoshi, T. Mikuriya and N.~Ishikawa are with the Faculty of Engineering, Yokohama National University, 240-8501 Kanagawa, Japan (e-mails: yukiyoshi-kein-wk@ynu.jp, mikuriya-taku-rw@ynu.jp, ishikawa-naoki-fr@ynu.ac.jp). This research was supported by the Japan Society for the Promotion of Science KAKENHI (Grant Numbers 22H01484 and 23K22755).}
\thanks{H.~S.~Rou and G.~T.~F.~Abreu are with the School of Computer Science and Engineering, Constructor University, Campus Ring 1, 28759, Bremen, Germany (e-mails: hrou@constructor.university, gabreu@constructor.university).}
\vspace{-4ex}}

\maketitle

\TPshowboxesfalse
\begin{textblock*}{\textwidth}(45pt,10pt)
\footnotesize
\centering
Accepted for publication in IEEE Transactions on Quantum Engineering. This is the author's version which has not been fully edited and content may change prior to final publication. Citation information: DOI 10.1109/TQE.2024.3450852
\end{textblock*}

\begin{abstract}
We propose new formulations of max-sum and max-min dispersion problems that enable solutions via the Grover adaptive search (GAS) quantum algorithm, offering quadratic speedup.
Dispersion problems are combinatorial optimization problems classified as NP-hard, which appear often in coding theory and wireless communications applications involving optimal codebook design.
In turn, GAS is a quantum exhaustive search algorithm that can be used to implement full-fledged maximum-likelihood optimal solutions.
In conventional naive formulations however, it is typical to rely on a binary vector spaces, resulting in search space sizes prohibitive even for GAS.
To circumvent this challenge, we instead formulate the search of optimal dispersion problem over Dicke states, an equal superposition of binary vectors with equal Hamming weights, which significantly reduces the search space leading to a simplification of the quantum circuit via the elimination of penalty terms.
Additionally, we propose a method to replace distance coefficients with their ranks, contributing to the reduction of the number of qubits.
Our analysis demonstrates that as a result of the proposed techniques a reduction in query complexity compared to the conventional GAS using Hadamard transform is achieved, enhancing the feasibility of the quantum-based solution of the dispersion problem.
\end{abstract}

\begin{IEEEkeywords}
Quantum Computing, Dispersion problem, Grover adaptive search (GAS), Dicke state, Codebook Design, Index Modulation, Multiple-access systems.
\end{IEEEkeywords}
\IEEEpeerreviewmaketitle


\vspace{-2ex}
\section{Introduction}
\IEEEPARstart{T}{he} dispersion problem \cite{chandrasekaran1981location, shier1977minmax, kuby1987programming, ghosh1996computational,erkut1990discrete,prokopyev2009equitable} is a combinatorial optimization problem with a wide range of applications in operations research, communications, computer science, and information theory.
A famous example is the facility location problem \cite{shier1977minmax, chandrasekaran1981location}, which concerns finding the optimal placement of $k$ grocery stores among $n$ possible locations within a city.
In such a scenario, it is preferable to maximize the distances between the stores, \ie, to increase the ``dispersion'' of the locations, so that each store can serve a different clientele, leading to maximum profits.
Another example is the deployment of ``mutually obnoxious'' facilities, such as nuclear power plants and oil storage tanks, which preferably should be located at a sufficiently far apart from one another so as to prevent an eventual isolated accident to propagate to the others in chain \cite{kuby1987programming}.

Similar problems are encountered in areas such as information retrieval and web search \cite{sydow2015approximation, cevallos2019improved}, social distancing \cite{kudela2020social} and optimal deployment of access points in wireless networks \cite{kim2012phub}.
In addition, although not often pointed out in the literature, the dispersion problem is essentially identical to the codebook design problem faced in coding theory and wireless communications \cite{plotkin1960binary,huang2022downlink}, which has been extensively studied independently from the dispersion problem.
To elaborate, in digital communication systems, information typically encoded as binary sequences must be mapped into complex-valued codewords $\mathbf{c}_i$ within a predefined codebook $\mathcal{C}$, such that the distances between codewords determine the overall performance of the system, because if two codewords are closely spaced, it becomes challenging for the receiver to distinguish these two codewords in the presence of noise.
In order to optimize system performance one must, therefore, design codebooks that maximize the dispersion of codewords.

The analytical model of the dispersion problem is defined as follows: given a set $\mathcal{P}$ of $n$ elements, and defining the distances $d_{i, j}$ between any pair of distinct elements $p_i$ and $p_j$, to be symmetric such that $d_{i, j} = d_{j, i}$, the objective is to find a subset $\mathcal{S} \subseteq \mathcal{P}$ of size $k$ such that a specified metric on the distances is maximized.
And although different types of dispersion problems exist, typically each as a result of a different distance metric of interest, the most common problems follows from one of the following two metrics: a) the summation of distances $\sum_{i < j} d_{i, j}$ for all pair of elements $p_i, p_j \in \mathcal{S}$, and b) the minimum distance $d_{\mathrm{min}} = \min_{i < j} \qty{d_{i, j} \mid p_i, p_j \in \mathcal{S}}$.

The first metric leads to the max-sum dispersion problem, which is equivalent to the max-avg (average) dispersion problem, addressed \eg in \cite{kuby1987programming,erkut1990discrete, ghosh1996computational}; while the second yields the max-min dispersion problem, also known as the $p$-dispersion problem studied \eg in \cite{shier1977minmax, chandrasekaran1981location, kuby1987programming, sayyady2016integer}.
Both these problems have been proven to be strongly NP-hard in general settings \cite{erkut1990discrete,kuo1993analyzing,ghosh1996computational}, motivating the development of various heuristic approaches \cite{kincaid1992good,hassin1997approximation, resende2010grasp,ravi1994heuristic} and a few polynomial-time algorithms designed under geometric assumptions \cite{ravi1994heuristic,wang1988study,akagi2018exact} to address their inherent difficulties.

Although Moore's Law -- which predicted in 1965 that the density of transistors on integrated circuits would double approximately every two years \cite{moore:1965} -- has held true for nearly half a century, it is widely anticipated that its end is near due to fundamental physical limitations, which motivates the research community to seek solace in the alternative of quantum computing \cite{choi2023quantum}.
And while demonstrating quantum supremacy remains challenging today\footnotemark, quantum computing is developing fast and anticipated to soon outperform classical computing in at least a selected number of tasks.
In particular, great effort in fault-tolerant quantum computers (FTQC) has been placed recently, which eliminate the effects of noise through error correction \cite{suzuki2022quantum} and have been shown to be advantageous over classical computers in terms of the query complexity required to solve certain problems.

\vspace{-1.75ex}
\footnotetext{Quantum annealing (QA) \cite{kadowaki1998quantum} and the quantum approximate optimization algorithm (QAOA) \cite{farhi2014quantum}, for instance, are two heuristic methods based on quantum mechanics which have already been demonstrated to work effectively on real devices, but which still cannot outperform classical computing in the presence of noise \cite{stilckfranca2021limitations}.}

\begin{table}[H]
\centering
\label{table:sym}
\caption{List of important mathematical symbols.}
\vspace{-1.5ex}
\begin{tabular}{lll}
$\{0, 1\}$ & & Finite field of order 2 \\
$\mathbb{R}$ & & Real numbers \\
$\mathbb{C}$ & & Complex numbers\\
$\mathbb{Z}$ & & Integers \\
$\mathrm{j}$ & $\in \mathbb{C}$ & Imaginary number\\
$e$ & $\in \mathbb {R}$ & Base of natural logarithm\\
$E(\cdot)$ & $\in \mathbb{R}$ & Objective function\\
$(\cdot)^\mathrm{H}$ & & Hermitian transpose\\
$n$ & $\in \mathbb {Z}$ & Number of binary variables\\
$m$ & $\in \mathbb{Z}$ & Number of qubits to encode $E(\cdot)$\\
$x$ & $\in \{0, 1\}$ & Binary variable \\
$\mathcal{X}$ & $\subseteq \{0, 1\}^n$ & A binary vector space\\
$\mathbf{x}$ & $\in \mathcal{X}$ & Binary variables \\
$t$ & $\in \mathbb{Z}$ & Number of solutions \\
$k$ & $\in \mathbb {Z}$ & Hamming weight\\
$\ket{D_{k}^{n}}$ &$\in \mathbb{C}^{2^n}$ & Dicke state\\
$\mathcal{D}$ & & A constant-weight binary vector space\\
$M$ & $\in \mathbb {Z}$ & Number of codewords\\
$y$ & $\in \mathbb{Z}$ & Threshold for $E(\cdot)$ \\
$L$ & $\in \mathbb{Z}$ & Number of applied Grover operators\\
$d_{i, j}$ & $\in \mathbb{R}$ & Distance between a pair of elements \\[0.5ex]
\hline
\end{tabular}
\vspace{-1.75ex}
\end{table}

But steady progress in quantum computing is also being made on algorithm design.
Consider for instance, the Grover's algorithm \cite{grover1996fast}, which finds one of $t$ solutions in an unordered database of $N$ elements with a query complexity of $\mathcal{O}(\sqrt{N/t})$\footnote{$\mathcal{O}(\cdot)$ denotes the big-O notation~\cite{knuth1976big}.}, in contrast to the classical computing search algorithm, which requires a query complexity of $\mathcal{O}(N/t)$.
A variation of this quantum algorithm, dubbed the Grover adaptive search (GAS) has recently been proposed \cite{durr1999quantum, bulger2003implementing, baritompa2005grover}, which can be interpreted as a quantum exhaustive search method over the search space, thus guaranteeing the global optimality of the obtained solution.

To cite another recent progress in this area, consider the fact that the search space in quantum search algorithms is typically prepared by a Hadamard transform, which produces the uniform superposition state of $\frac{1}{\sqrt{2^n}}\sum_{i=0}^{2^n-1} \Ket{i}$.
In certain problems, however, there may be only a subset $\{\ket{0}, \cdots, \ket{2^n - 1}\}$ of feasible solutions, such that the effective search space is sparse -- namely, consisting of of $N < 2^n$ feasible solutions -- which cannot be exactly prepared by the Hadamard transform.
In some studies \cite{ishikawa2021quantum, yukiyoshi2022quantum, sano2023qubit, sano2024accelerating}, this limitation is addressed by introducing a penalty term in the objective function that limits the search space.
Meanwhile, in \cite{sato2024circuit}, Grover's algorithm is used to prepare the exact search space of the traveling salesman problem.
In both these approaches, however, the query complexity is of order $\mathcal{O}(\sqrt{2^n/t})$, failing to achieve the full quadratic speedup of $\mathcal{O}(\sqrt{N/t})$, such that a loss in quantum speedup may result in cases when $\mathcal{O}(\sqrt{2^n/t}) > \mathcal{O}(N/t)$.

A promising approach for preparing exact search spaces for quantum search algorithms is the application of Dicke states $\ket{D_{k}^{n}}$ \cite{dicke1954coherence}.
A Dicke state is a higher-dimensional highly-entangled completely symmetric quantum superposition state, in which all $n$ qubits have an equal-weight $k$.
For example, a three-qubit Dicke state with a Hamming weight of two is expressed as $\ket{D_{2}^{3}} = (\ket{110} + \ket{011} + \ket{101})/\sqrt{3}$.

The primary application of Dicke states is to facilitate the superposition of all feasible solutions in combinatorial optimization problems such as the maximum $k$-vertex cover problem \cite{cook2020quantum} and the $k$-densest subgraph problem \cite{childs2002finding}.
But although the utilization of Dicke states has been well studied in the context of QAOA \cite{hadfield2019quantum, cook2020quantum, bartschi2020grover, golden2023numerical, yoshioka2023fermionic}, achieving a quantum advantage with QAOA under realistic assumptions is considered challenging \cite{stilckfranca2021limitations}. 

In turn, the application of Dicke states to FTQC schemes, including quantum search algorithms, has (to the best of our knowledge) not yet been explored.
Against this background, we formulate the dispersion problem as binary optimization problem that are suitable for GAS assuming FTQC.
The major contributions of the article are as follows:
\begin{enumerate}
\item We propose new binary formulation methods for the max-sum and max-min\footnotemark~dispersion problems, illustrating their applications to codebook design problems in wireless communication, thus expanding beyond the traditional whelm of operations research.
\item In order to circumvent the issue of search space expansion due to the Hadamard transform faced in original GAS algorithm, we replace the binary vector space with the constant-weight binary vector space using the Dicke state representation, resulting in scheme with full quadratic speedup.
To the best of our knowledge, the application of Dicke states to the FTQC algorithm is the first in the literature.
\item Finally, we propose a method to replace the distance coefficients in the objective function with their ranks, which further contributes to the reduction in the number of qubits required to implement the algorithm.
\end{enumerate}

\footnotetext{The max-min dispersion problem is very similar to the $k$-independent set problem \cite{akagi2018exact}, such that our contributions may also be extended to the quantum speedup of the latter problem.}

The remainder of this paper is as follows.
In Section \ref{sec:codebook}, we revisit three key problems in wireless communications and coding theory, establishing their relationship with the fundamental dispersion problem.
In Section \ref{sec:qs}, conventional quantum search algorithms, including GAS, are reviewed, followed by the introduction of a Dicke state-based formulation of GAS in Section \ref{sec:dicke}.
In Section \ref{sec:dispersion_problem}, we propose new quantum speed-up formulations for the dispersion problems.
Finally, the performance advantages of the proposed methods are justified in Section \ref{sec:performance} under a general setup, before conclusions are drawn in Section \ref{sec:conc}.

Table~\ref{table:sym} summarizes the important mathematical symbols used in this paper. Italicized symbols represent scalar values, while bold symbols represent vectors and matrices. We use zero-based indexing throughout this paper.

\section{Codebook Design and Dispersion Problems}
\label{sec:codebook}
In the design of modern communications systems, one is often required to obtain $K$ distinct solutions to a given problem, while simultaneously ensuring maximum mutual distances among the latter.
And while in some cases the solution space is continuous and unrestricted, such that the problem can be solved via iterative classical continuous optimization techniques \cite{Medra_TWC14, Zhang_TC17, Ghanem_CL23}, in other cases the solutions must be found within a discrete and finite domain, which renders the
continuous-reduced solutions suboptimal \cite{simon1989continuous,simon1990approximate}.
In such discrete and finite cases, the sought after solutions can be generally referred to as ``codewords'', such the problem of finding $K$ unique codewords within the space of given $N$ valid candidates can be seen as a codebook design problem, directly related to the classic dispersion problem \cite{plotkin1960binary,huang2022downlink}.

A fundamental challenge of such codebook design problems is that depending on the space $N$ of viable codewords, and the cardinality $K$ of the desired codebook, the total number $\binom{N}{K}$ of possible codebooks can be large enough to be prohibitive for an exhaustive (optimal) search to be performed with conventional computers, thus motivating the quantum computing solution to be introduced in the sequel.
In particular, we discuss in the following a few prominent codebook design problems of interest in communications systems, including binary codes, index modulation, and multiple access, casting each of these cases into the context of the fundamental dispersion problem addressed more generally in Section \ref{sec:dispersion_problem}.

\vspace{-2ex}
\subsection{Codebook Design for Binary Codes}

Let $\mathcal{B} \triangleq \{\mathbf{b}_n\}_{n=1}^{N} \in \{0,1\}^{L}$ be the set of all possible binary vectors of length $L$, where $|\mathcal{B}| \triangleq N = 2^L$.
In current literature on the design of unrestricted binary codes \cite{johnson1971upper,pang2023new}, it is typical to search for a codebook $\mathcal{C}$ containing $K$ distinct codewords from $\mathcal{B}$, satisfying the condition $d_{i,j} \geq D, ~\forall i\neq j$, where $d_{i,j}$ is the Hamming distance between the pair of codewords $\mathbf{c}_i$ and $\mathbf{c}_j$ in $\mathcal{B}$, and $D$ is a sufficient (or threshold) minimum distance.
Within such an approach, the search for a suitable codebook $\mathcal{C}$ depends therefore both on the minimum distance $D$ and the codebook size\footnote{The maximum possible size of an unrestricted binary codebook is known to be a function of $L$ and $D$ \cite{johnson1971upper, pang2023new}.} $K$.

It is clear, however, that such an approach is sub-optimum compared to the full max-min dispersion problem, in which no threshold minimum distance $D$ is pre-defined, but rather
the optimum codebook is the one whose minimum Hamming distance is the largest among all minimum Hamming distances of all possible codebooks of cardinality $K$.
In other words, a brute-force maximum-likelihood (ML) solution to the unrestricted binary codebook design problem requires a search over $\binom{N}{K} = \binom{2^L}{K}$ codebooks.

The same is true for binary constant weight codes \cite{brouwer1990new,ostergardTIT10,yukiyoshi2022quantum}, which are a special case of the above in which an additional constant weight constraint is added to ensure that all codewords in $\mathcal{C}$ to have exactly $W$ non-zero elements.
In this case, the solution requires the selection of $K$ codewords from a specific subset of binary vectors with weight $W$, \textit{i.e.,} $\mathcal{B}^{W} \triangleq \{\mathbf{b}_q\}_{q=1}^{Q} \subset \{0,1\}^{L}$, where $Q \triangleq \binom{L}{W}$.
This consequently implies the construction and evaluation of corresponding minimum Hamming distances of all $\binom{Q}{K} = \binom{\binom{L}{W}}{K}$ valid codebooks, in a similar manner to the unrestricted binary code.
In other words, and in short, both the unrestricted and binary constant weight codebook design problems are instances of the max-min dispersion problem to be addressed frontally  in Section \ref{sec:max-min-DP}.

With regards to this discussion, it is worth mentioning that a GAS-based quantum algorithm to finding binary constant weight codes was proposed in \cite{yukiyoshi2022quantum}.
The method thereby does, however, suffer from an increased query complexity due to the presence of the redundant search space resulting from the utilization of the Hadamard transform.
This limitation will be lifted here by applying Dicke states.

\vspace{-1ex}
\subsection{Codebook Design for Index Modulation Schemes}

Index modulation (IM) schemes \cite{ishikawa2019imtoolkit,ishikawa201850,Basar_Access17}, including notable instances such as spatial modulation (SM) \cite{mesleh2008spatial,direnzo2014spatial,an2022achievable}, are characterized by a unique transmission technique in which, at each transmission instance, only $W$ out of totally available $L$ resources are employed in order to convey information, encoded in the set of indices of the selected resources\footnote{Of course, information can also be encoded in the actually transmitted signals. For example, in SM-based schemes, complex symbols are modulated within each resource block. Such approaches can, however, can be decoupled into an equivalent purely IM formulations \cite{Rou_CAMSAP23}.}.
This characteristic is identical to the structure of binary constant weight codes, except for the fact that in IM, since binary information must be conveyed by the selection of the specific codeword from the codebook $\mathcal{C}$, the codebook size must be limited to a power of two.
It follows that each codebook of an IM scheme with $K$ codewords of length $L$ must be selected out of $Q^* \triangleq 2^{\lfloor\log_2{\binom{L}{W}}\rfloor}$ viable codewords, each encoding $\lfloor\log_2{\binom{L}{W}}\rfloor$ bits.
And since there are $\binom{Q}{Q^*}$ as many distinct choices\footnote{The set of viable codewords to design codebooks can be selected under different criteria. A classical linear combinatorial order approach was employed in \cite{Frenger_TC99, mesleh2008spatial}, while a diversity-optimal selections was proposed in \cite{Rou_TWC22}, and a lexicographical order given the CSIT was used in \cite{dang2018lexicographic}.} of viable codeword sets, there are a total of $\binom{Q}{Q^*}\cdot\binom{Q^*}{K}$ possible codebooks to be searched.

Although IM codebook design problems often use Euclidean distance as the pairwise distance metric\footnote{The rationale behind such approximation is that BER is dominated by most likely detection error events, which in turn are associated with codewords with shortest Euclidean distance at high SNRs. The argument does not hold, however, at low SNRs and therefore (although widely adopted) is not generally correct.} \cite{agrawal2000generalized,huang2022downlink}, under the assumption that the codebook is designed to optimize  communications performance in terms of bit error rates (BERs), binary Hamming distances should actually be used also for IM schemes \cite{ishikawa2021quantum,su2020codebook,han2021design}, such that all in all the problem is also equivalent to the max-min dispersion problem discussed in Section \ref{sec:max-min-DP}.

\vspace{-1ex}
\subsection{Codebook Design for Hybrid Beamforming}

Consider a multiple-input multiple-output (MIMO) multiple access scheme where a central base station (BS) equipped by an antenna array simultaneously serves $K$ user equipment (UEs), each also equipped with an antenna array, possibly of different sizes.
In such cases, it is typical that each UE transmits multiple streams of data, over a common resource interface, \eg\!\!, time slots and frequency carriers \cite{koutsopoulos2008impact,taherzadeh2014scma}.
In order to minimize multiuser interference in such multi-access schemes, each UE employs a unique beamforming matrix (aka precoder), known at the base station, and designed to separate the signals of different UEs, as received by the BS by corresponding receive beamformers (aka combiners) \cite{ding2015application}.

In principle, the design of such precoding and combining matrices can be achieved via optimization techniques over continuous spaces \cite{gershman2010convex,song2017common}, yielding optimum solutions consisting of complex-valued coefficients corresponding to the phase shifts and amplification gains to be applied at each antenna.
Recent trends towards systems operating high frequencies in the millimeter wave (mmWave) and terahertz (THz) bands impose, however, significant challenges onto the feasibility of such ``fully digital'' approaches, motivating the investigation of more cost-effective hybrid designs \cite{ahmed2018survey, zhu2019millimeter} in which the phase-shifts and amplification gains at the radio-frequency (RF) bands are restricted to a prescribed discrete and finite set of values\footnote{A similar situation emerges also in the design of reconfigurable intelligent surfaces (RIS) \cite{Ghanem_CL23,di2020hybrid,chen2017hybrid}}.

Without going into further details, as literature on the topic is rather vast \cite{nikopour2013sparse, xiao2018capacity,hong2022hybrid,alavi2018beamforming}, the optimum codebook design problem for hybrid beamforming as an enabler of MIMO multiple access systems requires selecting one combiner and $K$ precoders out of corresponding sets of $1 + N_k, k=\{1,\cdots,K\}$ valid candidates.
Aggregating the precoder spaces local to each UE and integrating the space of the combiner associated with the BS, it is clear that the problem can also be formulated as one -- somewhat expanded -- version of the max-min dispersion problem addressed in Section \ref{sec:max-min-DP}.

\vspace{-1ex}
\section{Quantum Search Algorithms}\label{sec:qs}

In this section, we review the quantum search and optimization algorithms that form the basis of our proposed approach.

\vspace{-2ex}
\subsection{Grover's Algorithm and Amplitude Amplification}

Grover's algorithm finds the desired solution in a database of unsorted $2^n$ elements \cite{grover1996fast} by amplifying the amplitude of the state corresponding to the desired solution, which is achieved by applying the Grover operator $\mathbf{G}$ to the uniform superposition state $\frac{1}{\sqrt{2^n}}\sum_{i=0}^{2^n-1} \Ket{i}$ a number of times.
As a result, the query complexity of Grover's algorithm, given by $\mathcal{O}(\sqrt{2^n})$, is fundamentally determined by the total number of oracle operators applied to quantum states.

The method was later extended to a generalized amplitude amplification framework \cite{brassard2002quantum}, which allows for multiple solutions to be searched, and for an almost arbitrary unitary transformation to be used instead of the Hadamard transform \cite{grover1998quantum}.
The extended version of Grover's algorithm enables the application to a wider range of the search spaces beyond the uniform superposition state.
Due to the augmented amplitude amplification mechanism, its query complexity is given by $\mathcal{O}(\sqrt{N/t})$, where $N$ is the size of the search space and $t$ is the number of solutions.

\vspace{-1ex}
\subsection{BBHT Algorithm}

In Grover's algorithm, the number of solutions $t$ must be known in advance to determine the optimal number of Grover operators to be applied.
In practice, however, this information is not available beforehand, which diminishes the practical usefulness of the method.
The Boyer-Brassard-H{\o}yer-Tapp (BBHT) algorithm \cite{boyer1998tight} seeks to mitigate this challenge  as briefly described below.

\begin{algorithm}[H]
\caption{BBHT Algorithm~\cite{boyer1998tight}.\label{alg:BBHT}}
\begin{algorithmic}[1]
\renewcommand{\algorithmicrequire}{\textbf{Input:}}
\renewcommand{\algorithmicensure}{\textbf{Output:}}
\REQUIRE $\lambda > 1$
\ENSURE $\mathbf{x}$

\STATE {Set $k_0 = 1$ and $i = 0$}.
\WHILE{Optimal solution not found}
\STATE\hspace{\algorithmicindent}{Randomly select the rotation count $L_i$ from the set $\{0, 1, ..., \lceil k_i-1 \rceil$\}}.
\STATE\hspace{\algorithmicindent}{Evaluate $\mathbf{G}^{L_i} \mathbf{A} \Ket{0}_{n}$ to obtain $\mathbf{x}$}.
\STATE\hspace{\algorithmicindent}{$k_{i+1}=\min{\{\lambda k_i,\sqrt{N}}\}$}.
\ENDWHILE
\end{algorithmic}
\end{algorithm}

In the BBHT algorithm, the number $L$ of Grover operators is treated as a random variable drawn from a uniform integer distribution $[0, k)$, where $\lambda$ is a constant value related to the increase rate of $k$, and the parameter $k$ is increased in each iteration by the rule of $k_{i+1} = \min\{\lambda k_i, \sqrt{N}\}$.
Following such an approach, an appropriate value of $L$ is first searched iteratively until the desired solution $\mathbf{x}$ is found through multiple measurements of the quantum states.
This iterative approach is shown to achieve the query complexity of $\mathcal{O}(\sqrt{N / t})$, even if the number of solutions is unknown.
The procedure is summarized in Algorithm \ref{alg:BBHT}.

\vspace{-1ex}
\subsection{Grover Adaptive Search for Binary Optimization}
\label{subsec:gas}

In short, GAS is a quantum algorithm that utilizes the BBHT algorithm to solve optimization problems, with the query complexity of $\mathcal{O}(\sqrt{N / t})$.
The GAS algorithm, which is summarized in Algorithm \ref{alg:GAS}, can be described as follows.
First, an initial threshold $y_0 = E(\mathbf{x}_0)$ is set by using a random input $\mathbf{x}_0 \in \mathcal{X}$, where $\mathcal{X}$ is a search space of the problem.
Then, the BBHT algorithm is used to search for a better solution $\mathbf{x}$ satisfying $E(\mathbf{x}) - y_i < 0 \Leftrightarrow E(\mathbf{x}) < y_i$, and then the threshold is updated via $y_{i+1} = E(\mathbf{x})$.
This process is repeated until a certain termination condition is met, and the optimal solution is finally obtained. 

The original GAS algorithm \cite{durr1999quantum,baritompa2005grover} assumes the existence of an efficient implementation of a black-box quantum oracle $\mathbf{O}$ that identifies the desired solutions.
A partial solution to this challenge was presented in \cite{gilliam2021grover}, where an efficient method was proposed to construct concrete quantum circuits for polynomial binary optimization problems in the form
\vspace{-0.5ex}
\begin{equation}
\begin{aligned}
\label{eq:PolynomialOptProb}
\min_{\mathbf{x}} \quad & E(\mathbf{x}) \\
\textrm{s.t.} \quad & \mathbf{x} \in \mathcal{X} \subseteq \{0, 1\}^n,
\end{aligned}
\end{equation}
where $\mathbf{x} = (x_0, \cdots\!, x_{n - 1})$ are binary variables, $\mathcal{X}$ represents the search space, and $E(\mathbf{x})$ denotes an arbitrary polynomial objective function.

The GAS circuit construction method described in \cite{gilliam2021grover} requires $n + m$ qubits, where $n$ is the number of binary variables and $m$ is the number of qubits to encode the objective function $E(\mathbf{x})$. Since $E(\mathbf{x})$ is expressed in the two's complement representation, $m$ must satisfy
\begin{equation}
-2^{m - 1} \leq \min_{\mathbf{x} \in \mathcal{X}} E(\mathbf{x}) \leq \max_{\mathbf{x} \in \mathcal{X}} E(\mathbf{x}) < 2^{m - 1}.
\end{equation}

\begin{figure*}[t]
\centering
\includegraphics[keepaspectratio,scale=0.7]{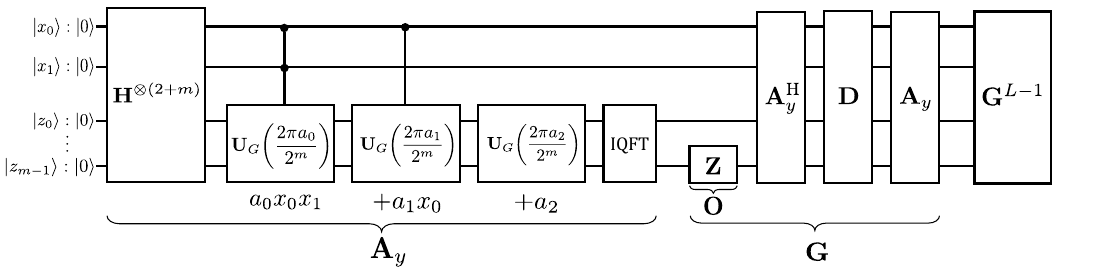}
\vspace{-2ex}
\caption{Quantum circuit of the GAS with the objective function $E(\mathbf{x}) - y = a_0x_0x_1 + a_1x_0 + a_2$.}
\label{fig:circuit1}
\vspace{1ex}
\hrule
\vspace{-3ex}
\end{figure*}

The objective is to construct a quantum circuit that prepares the quantum state $\mathbf{G}^{L} \mathbf{A}_{y} \ket{0}_{n+m}$, where $\mathbf{A}_{y}$ is the state preparation operator that encodes the values of $E(\mathbf{x}) - y$ into a quantum superposition for arbitrary inputs $\mathbf{x}$.
\newpage

For example, Fig. \ref{fig:circuit1} illustrates a quantum circuit constructed to represent the objective $E(\mathbf{x}) - y = a_0x_0x_1 + a_1x_0 + a_2$, where $a_2$ is a constant incorporating the constant terms in $E(\mathbf{x})$ and $-y$.
Each polynomial term $a_0x_0x_1$, $a_1x_0$, and $a_2$ is represented by a corresponding quantum gate $\mathbf{U}_{G}(\theta)$.
The construction involves the following steps.

\noindent {\bf 1) Uniform superposition:} Applying the Hadamard gates $\mathbf{H}^{\otimes (n+m)}$ to the initial state $\Ket{0}_{n+m}$ transforms it into a uniform superposition state.
This transition is expressed as~\cite{gilliam2021grover}
\begin{eqnarray}
\label{eq:AyH}
\Ket{0}_{n+m}\; \xrightarrow{\mathbf{H}^{\otimes (n+m)}} \hspace{-3ex}&& \frac{1}{\sqrt{2^{n+m}}} \sum_{i=0}^{2^{n+m}-1} \Ket{i}_{n+m} =\nonumber\\
&&\frac{1}{\sqrt{2^{n+m}}} \sum_{\mathbf{x} \in \{0, 1\}^{n}} \sum_{i=0}^{2^{m}-1} \Ket{\mathbf{x}}_{n} \Ket{i}_{m},
\end{eqnarray}
where, using the tensor product $\otimes$, the Hadamard gates $\mathbf{H}^{\otimes (n+m)}$ can be descreibed as
\begin{equation}
\mathbf{H}^{\otimes (n+m)} = \underbrace{\mathbf{H} \otimes \mathbf{H} \otimes \cdots \otimes \mathbf{H}}_{n+m}
\end{equation}
with
\begin{equation}
\mathbf{H} \triangleq \frac{1}{\sqrt{2}}\mqty[1 & 1 \\ 1 & -1].
\end{equation}

\noindent {\bf 2) Thresholding:} Each term of the difference $E(\mathbf{x}) - y$ comparing the objective function $E(\mathbf{x})$ with the threshold $y$ corresponds to a specific unitary gate $\mathbf{U}_{G}(\theta)$, which acts on the $m$ qubits and rotates the phase of quantum states. Given a coefficient of the term $a$, the phase angle is given by $\theta = \frac{2 \pi a}{2^{m}}$. The unitary gate is then defined as
\begin{equation}
\mathbf{U}_{G}(\theta) \triangleq \underbrace{\mathbf{R}(2^{m - 1}\theta) \otimes \cdots \otimes \mathbf{R}(2^{0}\theta)}_{m},
\end{equation}
where the phase gate $\mathbf{R}(\theta)$ is defined as
\begin{equation}
\mathbf{R}(\theta) \triangleq \mqty[1 & 0 \\ 0 & e^{\mathrm{j}\theta}].
\end{equation}

For a state $\mathbf{x}'$, where $\theta' = 2 \pi \qty(E(\mathbf{x}') - y) / 2^{m}$, the application of $\mathbf{U}_{G}(\theta')$ yields the transition \cite{gilliam2021grover}
\begin{equation}
\frac{1}{\sqrt{2^{m}}} \sum_{i=0}^{2^{m}-1} \Ket{i}_m \xrightarrow{\mathbf{U}_{G}\qty(\theta')}
\frac{1}{\sqrt{2^m}}\sum_{i=0}^{2^{m} - 1}e^{\mathrm{j}i\theta'} \Ket{i}_m.
\label{eq:AyUG}
\end{equation}

The interaction between binary variables is expressed by controlled gates $\mathbf{U}_{G}(\theta)$, as shown in Fig.~\ref{fig:circuit1}.

\noindent {\bf 3) Inverse Quantum Fourier Transform (IQFT):} The IQFT \cite{shor1997polynomialtime} acts on the lower $m$ qubits, yielding the transition \cite{gilliam2021grover}
\begin{equation}
\frac{1}{\sqrt{2^m}}\sum_{i=0}^{2^m - 1}e^{\mathrm{j}i\theta'} \Ket{i}_m  \xrightarrow{\mathrm{IQFT}} \frac{1}{\sqrt{2^m}}\Ket{E(\mathbf{x}_0) - y}_m.
\end{equation}

\begin{algorithm}[H]
\caption{Grover adaptive search (GAS)~\cite{gilliam2021grover,norimoto2023quantum}\label{alg:GAS}}
\begin{algorithmic}[1]
\renewcommand{\algorithmicrequire}{\textbf{Input:}}
\renewcommand{\algorithmicensure}{\textbf{Output:}}
\REQUIRE $E:\mathcal{X}\rightarrow\mathbb{R}, \lambda > 1$
\ENSURE $\mathbf{x}$
\STATE {Uniformly sample $\mathbf{x}_0 \in \mathcal{X}$ and set $y_0=E(\mathbf{x}_0)$}.

\STATE {Set $k = 1$ and $i = 0$}.

\REPEAT
\STATE\hspace{\algorithmicindent}{Randomly select the rotation count $L_i$ from the set $\{0, 1, ..., \lceil k-1 \rceil$\}}.
\STATE\hspace{\algorithmicindent}{Evaluate $\mathbf{G}^{L_i} \mathbf{A}_{y_i} \Ket{0}_{n + m}$ to obtain $\mathbf{x}$}.
\STATE\hspace{\algorithmicindent}{Evaluate $y = E(\mathbf{x})$}.
\hspace{\algorithmicindent}\IF{$y<y_i$}
\STATE\hspace{\algorithmicindent}{$\mathbf{x}_{i+1}=\mathbf{x}, y_{i+1}=y,$ and $k=1$}.
\hspace{\algorithmicindent}\ELSE{\STATE\hspace{\algorithmicindent}{$\mathbf{x}_{i+1}=\mathbf{x}_i, y_{i+1}=y_i,$ and $k=\min{\{\lambda k,\sqrt{N}}\}$}}.
\ENDIF
\STATE{$i=i+1$}.
\UNTIL{a termination condition is met}.
\end{algorithmic}
\end{algorithm}

The three steps above are collectively referred to as the \emph{state preparation operator} $\mathbf{A}_{y}$, which encodes the value $E(\mathbf{x})\! -\! y$ into $m$ qubits for arbitrary $\mathbf{x} \in \{0, 1\}^{n}$, satisfying \cite{gilliam2021grover}
\begin{equation}
\label{eq:Ay}
\mathbf{A}_{y} \Ket{0}_{n+m} = \frac{1}{\sqrt{2^n}}\sum_{\mathbf{x} \in \{0, 1\}^n} \Ket{\mathbf{x}}_{n}\Ket{E(\mathbf{x})-y}_{m}.
\end{equation}

\noindent {\bf 4) Identification:} Finally, in order to amplify the desired states, the Grover operator $\mathbf{G} = \mathbf{A}_{y}\mathbf{F}\mathbf{A}_{y}^H\mathbf{O}$ is applied $L$ times, where $\mathbf{F}$ denotes the Grover diffusion operator \cite{grover1996fast}.
The oracle $\mathbf{O}$ identifies the desired states satisfying $E(\mathbf{x}) - y < 0$. Given that we utilize the two's complement representation, such states can be identified by a single Pauli-Z gate
\begin{equation}
\mathbf{Z} \triangleq \mqty[1 & \;\;0 \\ 0 & -1],
\end{equation}
acting on the most significant qubit of $m$ qubits.

While the coefficients of the objective function are restricted to integers in \cite{gilliam2021grover}, it has been shown in \cite{norimoto2023quantum} that the quantum circuit can be extended to accommodate real-valued coefficients at the expense of slightly increased query complexity.
This performance penalty can be circumvented with a sufficiently large number of $m$ qubits representing real-valued coefficients by approximated integers.

\section{Grover Adaptive Search with Dicke States}
\label{sec:dicke}

It is typical for quantum search algorithms \cite{ishikawa2021quantum, yukiyoshi2022quantum, norimoto2023quantum, huang2023optimizationa,ye2019quantum,yoder2014fixedpoint} to assume that the initial state is prepared by the Hadamard transform, which produces the uniform superposition of $\frac{1}{\sqrt{2^n}}\sum_{i=0}^{2^n-1} \ket{i}$, while the search space may be a sparse subset of $\{\ket{0}, \cdots, \ket{2^n - 1}\}$.

Consider, for instance, the dispersion problem of selecting a subset of size $2$ from $3$ elements. If we represent the presence of the $i$-th element in the subset by the $i$-th qubit being $1$, the resulting search space is $\{\ket{110}, \ket{101}, \ket{011}\}$, rather than the entire superposition space $\{\ket{000}, \cdots, \ket{111}\}$.
In this section, we address this limitation by replacing the Hadamard transform $\mathbf{H}^{\otimes n}$ in the GAS circuit with the Dicke state preparation operator $\mathbf{U}_{D}^{n, k}$.
It has been shown that the Hadamard transform can be replaced by any unitary operation that spans the search space \cite{grover1998quantum}, thereby justifying this substitution.

\vspace{-1ex}
\subsection{Dicke State Preparation via B{\"a}rtschi's Method}

Literature exists on methods to construct quantum circuits to prepare Dicke states $\ket{D_{k}^{n}}$ \cite{childs2002finding, bartschi2019deterministic, mukherjee2020preparing, aktar2022divideandconquer, bartschi2022shortdepth}.
One of the most recent, hereafter adopted as the state-of-the-art (SotA) method, was proposed by B{\"a}rtschi in \cite{bartschi2022shortdepth}, which has a circuit depth of $\mathcal{O}(k \log (n/k))$, a total of $\mathcal{O}(nk)$ CNOT gates, and no ancilla qubits.
It is claimed in \cite{bartschi2022shortdepth} that the method achieves an optimal circuit depth, up to constant factors.
A detailed description of the method is beyond the scope of this paper, but a brief overview of the basic concepts is offered in the sequel for the convenience of the reader.
We assume an all-to-all connectivity for the qubit topology and use little-endian notation to represent quantum states.

B{\"a}rtschi's method \cite{bartschi2022shortdepth} consists of two components: weight distribution block $\mathbf{WDB}_{k}^{n, m}$, and the Dicke state unitary operator $\mathbf{U}_{k}^{n}$, which satisfies \cite{bartschi2019deterministic}
\begin{align}
\label{eq:dicke_unitary}
\ket{1^l0^{n-l}}_{n} \xrightarrow{\mathbf{U}_{k}^{n}} \ket{D_{l}^{n}}_{n},
\end{align}
for all integers $l$ with $1 \leq l \leq k$.

The construction method of the Dicke state unitary in \cite{bartschi2019deterministic} requires a circuit depth of $\mathcal{O}(n)$ and $\mathcal{O}(nk)$ two-qubit gates.
By applying the Dicke state unitary to the state $\ket{1^{k}0^{n-k}}_{n}$, we can prepare the Dicke state, however it requires a circuit depth of $\mathcal{O}(n)$.
In \cite{bartschi2022shortdepth}, B{\"a}rtschi \etal reduced the circuit depth to $\mathcal{O}(k \log (n/k))$ by introducing the weight distribution block $\mathbf{WDB}_{k}^{n, m}$, defined as an operator that satisfies \cite{bartschi2022shortdepth}
\begin{eqnarray}
\ket{1^l0^{n-l}}_{n} \xrightarrow{\mathbf{WDB}_{k}^{n, m}}&&\\
&&\hspace{-20ex}\frac{1}{\sqrt{\tbinom{n}{l}}} \sum_{i=0}^{l} \sqrt{\tbinom{m}{i}\tbinom{n-m}{l-i}}\ket{1^{l-i}0^{n-m+i-l}}_{n-m}\ket{1^{i}0^{m-i}}_{m},\nonumber
\end{eqnarray}
for all integers $l$ with $1 \leq l \leq k$.

As its name indicates, the block $\mathbf{WDB}_{k}^{n, m}$ distributes the input weights $l \leq k$ across two sets of qubits: one containing $n-m$ qubits and the other containing $m$ qubits.
Interestingly, the Dicke state $\ket{D_{k}^{n}}$ can be prepared by first applying $\mathbf{WDB}_{k}^{n, m}$ to the quantum state $\ket{1^{k}0^{n-k}}_{n}$ and then applying the Dicke state unitary operators $\mathbf{U}_{k}^{n-m}$ and $\mathbf{U}_{k}^{m}$ to each set of $n-m$ and $m$ qubits, respectively.
This property enables the recursive application of $\mathbf{WDB}_{k}^{n, m}$ until the size of each qubit group becomes equal to, or smaller than, the weight $k$, analogous to the construction of a binary tree.
Finally, the Dicke state $\ket{D_{k}^{n}}$ is prepared by applying the respective Dicke state unitaries $\mathbf{U}_{i}^{i}$ to all qubit groups, satisfying $i \leq k$.

\begin{figure}[H]
\includegraphics[width=\columnwidth]{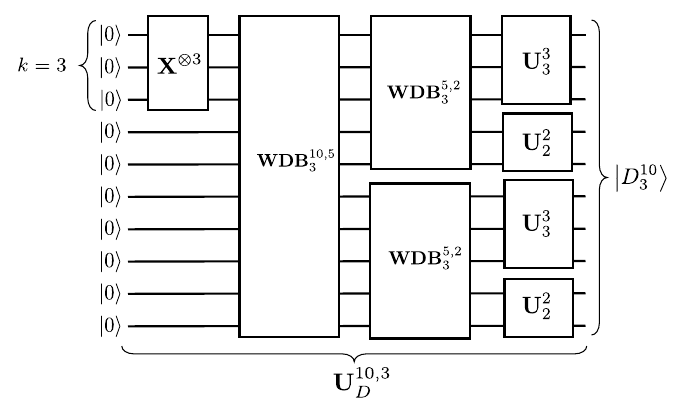}
\vspace{-4ex}
\caption{Quantum circuit $\mathbf{U}_{D}^{10, 3}$ that prepares the Dicke state $\ket{D_{3}^{10}}$.}
\label{fig:circuit2}
\vspace{-3ex}
\end{figure}

Fig.~\ref{fig:circuit2} illustrates a circuit corresponding to the Dicke state preparation operator $\mathbf{U}_{D}^{10, 3}$ that prepares $\ket{D_{3}^{10}}$, where $\mathbf{X}$ in the figure denotes a Pauli-X gate defined by
\newpage

\begin{equation}
\mathbf{X} = \mqty[0 & 1 \\ 1 & 0],
\end{equation}
which flips the basis states $\ket{0}$ and $\ket{1}$.

Slightly differently from the definition of the Dicke state unitary operator in \eqref{eq:dicke_unitary}, the Dicke state preparation operator $\mathbf{U}_{D}^{n, k}$ is defined as an operator that satisfies
\begin{equation}
\ket{0}_{n} \xrightarrow{\mathbf{U}_{D}^{n, k}} \ket{D_{k}^{n}}_{n}.
\end{equation}

The construction method of the weight distribution block in \cite{bartschi2022shortdepth} requires a circuit depth of $\mathcal{O}(k)$ and $\mathcal{O}(k^2)$ two-qubit gates. Given that the depth of the recursion is at most $\mathcal{O}(\log(n/k))$ due to the logarithmic height of the binary tree, the total depth of the preparation circuit is given by $\mathcal{O}(k\log(n/k))$.

\subsection{Integrating Dicke States into GAS}
\label{subsec:gas_with_dicke}

In order to integrate Dicke states into the GAS algorithm, the uniform superposition state prepared by the state preparation operator $\mathbf{A}_y$ from the conventional uniform superposition $\frac{1}{\sqrt{2^n}}\sum_{i=0}^{2^n-1}\ket{i}$ must be replaced with the Dicke state $\ket{D_{k}^{n}}$.
Thus, the step 1) of the circuit construction method in Section \ref{subsec:gas} is revised as follows:

\noindent {\bf 1) Uniform superposition:} Applying the Dicke state preparation operator $\mathbf{U}_{D}^{n, k}$ to $n$ qubits in the initial state $\Ket{0}_{n+m}$ transforms it into a Dicke state, with the corresponding transition described by
\begin{equation}
\Ket{0}_{n+m} \xrightarrow{\mathbf{U}_{D}^{n, k}} \ket{D_{k}^{n}} \Ket{0}_{m} = \frac{1}{\sqrt{\tbinom{n}{k}}} \sum_{\mathbf{x} \in \mathcal{D}} \Ket{\mathbf{x}}_{n} \Ket{0}_{m},
\end{equation}
where the reduced search space $\mathcal{D}$ is a set of binary variables $\mathbf{x}$ with a Hamming weight of $k$, \ie, $|\mathcal{D}| = \tbinom{n}{k}$.

Then, as with the original GAS, Hadamard gates are applied to $m$ qubits, yielding the transition
\begin{equation}
\frac{1}{\sqrt{\tbinom{n}{k}}} \sum_{\mathbf{x} \in \mathcal{D}} \Ket{\mathbf{x}}_{n} \Ket{0}_{m}\xrightarrow{\mathbf{H}^{\otimes m}} \frac{1}{\sqrt{\tbinom{n}{k}2^m}} \sum_{\mathbf{x} \in \mathcal{D}} \sum_{i=0}^{2^{m}-1} \Ket{\mathbf{x}}_{n} \Ket{i}_{m}.
\end{equation}

\begin{figure*}[t]
\centering
\includegraphics[keepaspectratio,scale=0.7]{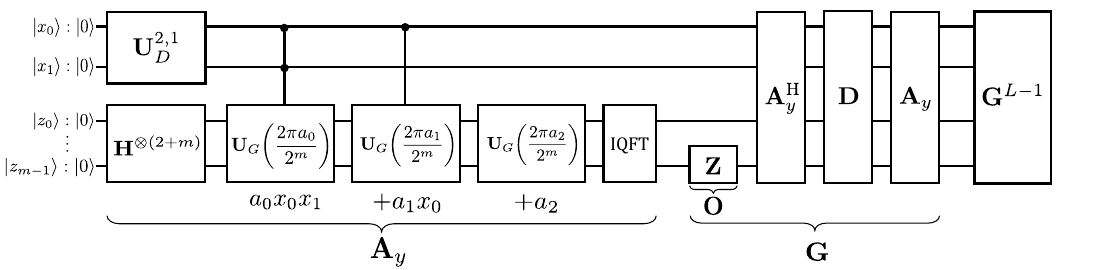}
\vspace{-2ex}
\caption{Quantum circuit of the GAS with Dicke state and the objective function $E(\mathbf{x}) - y = a_0x_0x_1 + a_1x_0 + a_2$, subject to the constraint $x_0 + x_1 = 1$.}
\label{fig:circuit3}
\vspace{1ex}
\hrule
\vspace{-3ex}
\end{figure*}

The remaining steps  are the same as those presented in Section \ref{subsec:gas}.
As a result, the transition yielded by the state preparation operator $\mathbf{A}_y$ is modified from \eqref{eq:Ay} to
\begin{equation}
\mathbf{A}_{y} \Ket{0}_{n+m} = \frac{1}{\sqrt{\tbinom{n}{k}}}\sum_{\mathbf{x} \in \mathcal{D}} \Ket{\mathbf{x}}_{n}\Ket{E(\mathbf{x})-y}_{m}.
\end{equation}

In contrast to Fig.~\ref{fig:circuit1}, the quantum circuit shown in Fig.~\ref{fig:circuit3} shows is constructed to represent the objective function $E(\mathbf{x}) - y = a_0x_0x_1 + a_1x_0 + a_2$, subject to the constraint $x_0 + x_1 = 1$.
The query complexity the GAS with and without Dicke states are given by $\mathcal{O}(\sqrt{\binom{n}{k}/t})$ and $\mathcal{O}(\sqrt{2^n/t})$, respectively.

\vspace{-1ex}
\subsection{Number of Quantum Gates}

Since the feasibility of a quantum algorithm is directly related to the number of required quantum gates, we analyze the number of quantum gates in the GAS algorithm with Dicke state preparation, focusing on the number of quantum gates required by the state preparation operator $\mathbf{A}_{y}$, which is the dominant component in the GAS circuit \cite{gilliam2021grover, norimoto2023quantum}.

The construction of the weight distribution block $\mathbf{WDB}_{k}^{n, m}$ in \cite{bartschi2022shortdepth} requires $\mathcal{O}(k^2)$ controlled-X (CX) gates, $\mathcal{O}(k^2)$ controlled-CX (CCX) gates, $\mathcal{O}(k)$ controlled-R (CR) gates, and $\mathcal{O}(k^2)$ controlled-CR (CCR) gates.
Since $\sum_{i=0}^{\mathcal{O}(\log(n/k))} 2^i = \mathcal{O}(n/k)$ weight distribution blocks are required, the total number of quantum gates needed in the recursive weight distribution steps are $\mathcal{O}(k^2 \cdot n/k) = \mathcal{O}(nk)$ for CX gates, $\mathcal{O}(nk)$ for CCX gates, $\mathcal{O}(k \cdot n/k) = \mathcal{O}(n)$ for CR gates, and $\mathcal{O}(nk)$ for CCR gates.
Similarly, the construction of the Dicke state unitary operator $\mathbf{U}_{k}^{k}$ requires $\mathcal{O}(k^2)$ CX gates, $\mathcal{O}(k)$ CR gates, and $\mathcal{O}(k^2)$ CCR gates
Given that $\mathcal{O}(n/k)$ Dicke state operators are required, the total number of quantum gates for the final step are $\mathcal{O}(k^2 \cdot n/k) = \mathcal{O}(nk)$ for CX gates, $\mathcal{O}(k \cdot n/k) = \mathcal{O}(n)$ for CR gates, and $\mathcal{O}(nk)$ for CCR gates.
Consequently, the total number of quantum gates required by the entire Dicke state preparation circuit $\mathbf{U}_{D}^{n, k}$ is $\mathcal{O}(nk + nk) = \mathcal{O}(nk)$ for CX gates, $\mathcal{O}(nk + 0) = \mathcal{O}(nk)$ for CCX gates, $\mathcal{O}(n + n) = \mathcal{O}(n)$ for CR gates, and $\mathcal{O}(nk + nk) = \mathcal{O}(nk)$ for CCR gates.

According to \cite{gilliam2021grover}, the original GAS circuit with Hadamard transform for quadratic objective functions requires $\mathcal{O}(n + m)$ Hadamard gates, $\mathcal{O}(m)$ R gates, $\mathcal{O}(nm)$ CX gates, $\mathcal{O}(nm)$ CR gates, and $\mathcal{O}(n^2m)$ CCR gates. Given that the GAS with Dicke state replaces $n$ Hadamard gates with the Dicke state preparation circuit $\mathbf{U}_{D}^{n, k}$, the GAS with Dicke state requires $\mathcal{O}(m)$ Hadamard gates, $\mathcal{O}(m)$ R gates, $\mathcal{O}(nm + nk) = \mathcal{O}(n(m + k))$ CX gates, $\mathcal{O}(nk)$ CCX gates, $\mathcal{O}(nm + n) = \mathcal{O}(nm)$ CR gates, and $\mathcal{O}(n^2m + nk) = \mathcal{O}(n^2m)$ CCR gates.
\vspace{-2ex}
\begin{table}[H]
\centering
\label{table:gate}
\caption{Number of quantum gates required by $\mathbf{A}_{y}$.}
\vspace{-2ex}
\begin{tabular}{lll}
\hline
Gate & Conv. GAS~\cite{gilliam2021grover} &  GAS with Dicke state\\
\hline
H & $\mathcal{O}(n + m)$ & $\mathcal{O}(m)$ \\
R & $\mathcal{O}(m)$ & $\mathcal{O}(m)$ \\
CX & $\mathcal{O}(nm)$ & $\mathcal{O}(n(m + k))$ \\
CCX & 0 & $\mathcal{O}(nk)$ \\
CR & $\mathcal{O}(n m)$ & $\mathcal{O}(nm)$ \\
CCR & $\mathcal{O}(n^2m)$ & $\mathcal{O}(n^2m)$ \\
\hline
\end{tabular}
\vspace{-3ex}
\end{table}

Given the analysis above, the number of quantum gates required to implement the state preparation operator $\mathbf{A}_{y}$ is summarized in Table \ref{table:gate}.

\section{Quantum Speedups for the Dispersion and Codebook Design Problems}
\label{sec:dispersion_problem}

Finally, we turn our attention to showing how the GAS algorithm with Dicke state preparation can be employed to accelerate the exhaustive search associated with the ML solution of general dispersion problems and, consequently, of the various codebook design problems related to the latter, as discussed in Section \ref{sec:codebook}.

We begin by formulating the max-sum and max-min dispersion problems as binary optimization problems, aiming to obtain a solution by GAS, and then introduce a distance compression technique that improves the feasibility of GAS using our formulation for the max-min dispersion problem.
Lastly, we review the time complexities of the current fastest classical algorithms for the the max-sum and max-min dispersion problems, and discuss the potential for quantum speedup.

\subsection{Max-Sum Dispersion Problem}
\label{subsec:max-sum-DP}

The max-sum dispersion problem can be formulated as the following polynomial binary optimization problem
\begin{equation}
\begin{aligned}
\min_{\mathbf{x}} \quad & E_1(\mathbf{x}) \\
\textrm{s.t.} \quad & \mathbf{x} \in \mathcal{X},
\end{aligned}
\end{equation}
with the objective function \cite{kuby1987programming}
\begin{equation}
\label{eq:objfun_maxsum}
E_1(\mathbf{x}) = -\!\!\!\! \underbrace{\sum_{i < j} d_{i, j} x_i x_j}_{\geq \tbinom{k}{2}(\max_{i < j} d_{i, j})}\!\!\!\! + \lambda_2 \qty(\sum_{i=0}^{n-1} x_i - k)^{\!\!2}\!.
\end{equation}

Here, $\mathbf{x} = (x_0, ~ \cdots, x_{n - 1})$ is a vector of binary variables representing whether an element $p_i$ is included in the subset $\mathcal{S}$ or not, while $\lambda_2$ is a penalty coefficient.
The second term in equation \eqref{eq:objfun_maxsum} is a penalty function that constrains the size of the subset $\mathcal{S}$ to $K$, and since the magnitude of the first term is lower-bounded as indicated, we can ensure that the constraint is satisfied by setting the penalty coefficient $\lambda_2$ equal to $\tbinom{k}{2}(\max_{i < j} d_{i, j})$.

\subsection{Max-Min Dispersion Problem}
\label{sec:max-min-DP}

The max-min dispersion problem can be formulated as the polynomial binary optimization problem
\begin{equation}
\begin{aligned}
\min_{\mathbf{x}} \quad & E_2(\mathbf{x}) \\
\textrm{s.t.} \quad & \mathbf{x} \in \mathcal{X},
\end{aligned}
\end{equation}
where we propose a novel objective function
\begin{equation}
\label{eq:objfun_maxmin}
E_2(\mathbf{x}) = \sum_{i < j} \qty(\frac{1}{d_{i, j}})^{\!\lambda_1}\!\!\!\! x_i x_j + \lambda_2 \qty(\sum_{i=0}^{n-1} x_i - k)^{\!\!2}\!,
\end{equation}
and the vector $\mathbf{x}$ and penalty coefficients $\lambda_1$ and $\lambda_2$ are similar to those in Subsection \ref{subsec:max-sum-DP}.

Notice that the first term in equation \eqref{eq:objfun_maxmin} is for finding a subset with maximized $d_{\mathrm{min}}$, while the second term is a penalty function that constrains the size of the subset $S$ to $k$.
The following bounding theorem holds for the objective function in equation \eqref{eq:objfun_maxmin}.

\begin{thm}
Assuming $d_{i, j} \geq 1$ for arbitrary pair of distinct indices $(i, j)$, the minimum distance $d_{\mathrm{min}}$ of the obtained subset from the optimal solution of \eqref{eq:objfun_maxmin} is maximized, if the penalty coefficient $\lambda_1$ satisfies
\begin{equation}
\label{eq:thm1}
\qty(\frac{1}{d_{i, j}})^{\!\!\lambda_1} > \binom{k}{2}\qty(\frac{1}{d_{i', j'}})^{\!\!\lambda_1},
\end{equation}
for any arbitrary pair of distances $d_{i, j}$ and $d_{i', j'}$ satisfying $d_{i, j} < d_{i', j'}$.
\end{thm}

\begin{proof}
Let $d_{\mathrm{min}}^{(1)}$ and $d_{\mathrm{min}}^{(2)}$ be the maximum and second minimum possible $d_{\mathrm{min}}$ among all the possible subsets $\mathcal{S}$, respectively.
The subset obtained from the optimal solution of equation \eqref{eq:objfun_maxmin} should maximize $d_{\mathrm{min}}$, \ie, $d_{\mathrm{min}} = d_{\mathrm{min}}^{(1)}$.
By ignoring the second term in the equation, without loss of generality, an upper bound for the optimal subsets with $d_{\mathrm{min}} = d_{\mathrm{min}}^{(1)}$ is obtained, which is given by
\begin{equation}
\label{eq:upper_bound}
E_2(\mathbf{x}) \leq \binom{k}{2} \qty(\frac{1}{d_{\mathrm{min}}^{(1)}})^{\lambda_1},
\end{equation}
while the lower bound of equation \eqref{eq:objfun_maxmin} for other non-optimal subsets with $d_{\mathrm{min}} \leq d_{\mathrm{min}}^{(2)}$ is given by
\begin{equation}
\label{eq:lower_bound}
E_2(\mathbf{x}) \geq \qty(\frac{1}{d_{\mathrm{min}}^{(2)}})^{\lambda_1}.
\end{equation}

When the condition \eqref{eq:thm1} is satisfied for arbitrary pair of distances $d_{i, j}$ and $d_{i', j'}$ satisfying $d_{i, j} < d_{i', j'}$, the upper bound for the optimal subsets \eqref{eq:upper_bound} is less than the lower bound for other non-optimal subsets \eqref{eq:lower_bound}. Consequently, the subset obtained from the optimal solution of equation \eqref{eq:objfun_maxmin} with the minimum objective function value satisfies $d_{\mathrm{min}} = d_{\mathrm{min}}^{1}$.
\end{proof}

Using Theorem~1, we can set the penalty coefficient $\lambda_1$ to
\begin{equation}
\label{eq:lambda_1}
\lambda_1 = \max_{d_{i, j} < d_{i', j'}} \frac{\log k + \log (k + 1) - \log 2} {\log (d_{i', j'}) - \log (d_{i, j})},
\end{equation}
which enables us to obtain the optimal solution by minimizing the objective function \eqref{eq:objfun_maxmin}.

Given that the first term in equation \eqref{eq:objfun_maxmin} is upper bounded by $\tbinom{k}{2}(1 / \min_{i < j} d_{i, j})^{\lambda_1}$, we can ensure that the constraint is satisfied by setting $\lambda_2 = \tbinom{k}{2}(1 / \min_{i < j} d_{i, j})^{\lambda_1}$.
There are, however, two potential problems with the above formulation for the max-min dispersion problem, which are addressed in the sequel.

First, what if there exists a distance $d_{i, j}$ such that $d_{i, j} < 1$, which violates the assumption of Theorem 1?
Second, the distance $d_{i, j}$ and the penalty coefficient $\lambda_1$, obtained from equation \eqref{eq:lambda_1}, can be large numbers, resulting in exponentially small coefficients in the objective function \eqref{eq:objfun_maxmin}.
This may require a large register size to express real-valued numbers during computation, possibly compromising the feasibility of the GAS algorithm.

Fortunately, these challenges can be overcome without additional burden by introducing a distance compression technique.
To that end, suffice it to observe that for the actual values of the distances $d_{i, j}$ do not matter to the solution of the max-min dispersion problem, as long as the relative size relationship between any pair of distances is maintained.
For example, adding or subtracting the same constant value to all distances $d_{i, j}$ does not affect the solution of the problem.

Let us therefore define a rank function $R(d_{i, j})$ that returns the size rank of a distance $d_{i, j}$ among all distances, starting from $0$.
Compressing all distances $d_{i, j}$ to $1 + R(d_{i, j}) \cdot \delta$, where $\delta$ is an arbitrarily positive constant, preserves the relative size relationship between any pair of distances.
For example, consider a distance matrix given by
\begin{equation}
(d_{i,j}) = \mqty
[- & 2 & 7 & 9\\
2 & - & 6 & 7\\
7 & 6 & - & 5\\
9 & 7 & 5 & - ],
\end{equation}
where $(i, j)$ element of the matrix represents the distance $d_{i, j}$. The rank function $R(d_{i, j})$ assigns ranks as follows: $R(2) = 0, R(5) = 1, R(6) = 2, R(7) = 3$ and $R(9) = 4$.

Thus, the compressed distance matrix is given by
\begin{equation}
(d_{i,j}') = \mqty
[- & 1 & 1 + 3\delta & 1 + 4\delta\\
1 & - & 1 + 2\delta & 1 + 3\delta\\
1 + 3\delta & 1 + 2\delta & - & 1 + \delta\\
1 + 4\delta & 1 + 3\delta & 1 + \delta & -].
\end{equation}

After applying distance compression, the minimum coefficient of the objective \eqref{eq:objfun_maxmin} becomes a function of the maximum rank $r_{max} = \max_{i<j} R(d_{i, j})$ and $\delta$, which is expressed as
\begin{equation}
f(r_{max}, \delta) = \qty(\frac{1}{1 + r_{max}\delta})^{\!\!\lambda_1}\!\!\!,
\end{equation}
where, according to equation \eqref{eq:lambda_1}, $\lambda_1$ is given by
\begin{equation}
\lambda_1 = \frac{\log k + \log (k + 1) - \log 2}{\log (1 + r_{max}\delta) - \log (1 + (r_{max} - 1)\delta)}.
\end{equation}

When $r_{max}$ is fixed, $f(r_{max}, \delta)$ is a monotonically decreasing function of $\delta$.
As the value of $\delta$ decreases, $f(r_{max}, \delta)$ increases, becoming easier to encode.
Therefore, $\delta$ should be set as small as possible.

The limit of $f(r_{max}, \delta)$ as $\delta$ approaches zero is given by
\begin{equation}
\label{eq:lim}
\lim_{\delta \to 0^+} f(r_{max}, \delta) = \exp\qty{-r_{max}\qty(\log k + \log (k + 1) - \log 2)},
\end{equation}
from which it follows that the required number of register qubits $m$ is estimated to be $\mathcal{O}(r_{max}\log k)$, since $m$ is proportional to the logarithm of the number expressed. 
\newpage

All in all, this technique of distance compression allows us to satisfy the assumptions of Theorem 1 and prevents the coefficients of the objective function \eqref{eq:objfun_maxmin} to vanish, thus enhancing the feasibility of GAS using the proposed formulation, especially when solved by a quantum computer with current strict resource limitations.
We also emphasize that if the distances $d_{i, j}$ are sorted, the complexity associated with determining $\lambda_1$ from \eqref{eq:lambda_1} and applying distance compression is $\mathcal{O}(n^2 \log n^2) = \mathcal{O}(n^2 \log n)$, which is negligible compared to the complexity of the problem itself.

\vspace{-2ex}
\subsection{Max-Sum/Min Dispersion Problem: Final Formulations}
\label{subsec:max-sum/min-DP}

Having described how the max-sum and the max-min dispersion problems can be cast as polynomial binary optimization problems and solved by applying GAS with Dicke states, we now concisely describe the final formulation of these problems, for the sake of clarity.
Straightforwardly, the max-sum dispersion problem can be formulated as
\begin{equation}
\begin{aligned}
\min_{\mathbf{x}} \quad & - \sum_{i < j} d_{i, j} x_i x_j \\
\textrm{s.t.} \quad & \mathbf{x} = (x_0, ~ \cdots, x_{n-1}) \in \mathcal{D},
\end{aligned}
\end{equation}
while the max-min dispersion problem can be formulated as
\begin{equation}
\begin{aligned}
\min_{\mathbf{x}} \quad & \sum_{i < j} \qty(\frac{1}{d_{i, j}})^{\lambda_1} x_i x_j \\
\textrm{s.t.} \quad & \mathbf{x} = (x_0, ~ \cdots, x_{n-1}) \in \mathcal{D},
\end{aligned}
\end{equation}
where we highlight that the removal of the second terms in the objective functions \eqref{eq:objfun_maxsum} and \eqref{eq:objfun_maxmin}, enabled by the arguments described in Subsections \ref{subsec:max-sum-DP} and \ref{sec:max-min-DP}, help reduce the number of register qubits $m$ and thus the number of quantum gates required for $\mathbf{A}_y$, such that both problems can be solved by the GAS algorithm with Dicke states proposed in Section \ref{subsec:gas_with_dicke}.

Notice also that thanks to this contribution, the optimized search conducted to solve both problems takes place within the search space $\mathcal{D}$ satisfying $\sum_{i} x_i = k$, yielding quadratic speedup in query complexity.
In other words, in Algorithm \ref{alg:GAS}, the search space $\mathcal{X}$ from which the initial uniform sampling of the random solution in GAS is drawn, is reduced to $\mathcal{D}$.

\vspace{-2ex}
\subsection{A Note on Complexity: Feasibility of Quantum Speedup}

The comparison of the complexities between classical and quantum algorithms is challenging in general due to the distinct concepts of time complexity in classical algorithms and query complexity in quantum algorithms.
Additionally, the execution time of a quantum algorithm is highly dependent on specific hardware implementation. Despite these challenges, understanding the complexities of both classical and quantum algorithms is informative when exploring the potential for quantum speedup.

While the time complexity of the exact algorithm for the max-sum dispersion problem is not often discussed in the literature, the $k$-clique problem can be reduced to the max-sum dispersion problem in polynomial time.
This reduction is employed during the proof of NP-hardness \cite{kuo1993analyzing}.
Consequently, the max-sum dispersion problem is considered to be ``equally or more difficult'' than the $k$-clique problem, such that the time complexity of the max-sum dispersion problem can be estimated to be at least as high as that of the best classical algorithm for the $k$-clique problem \cite{nesetril1985complexity}, namely $\mathcal{O}(n^{\omega k/3})$, where $\omega$ is the matrix multiplication exponent.

Since the best bound currently known on the complexity exponent of matrix multiplication is $\omega \leq 2.371552$~\cite{williams2024new}, this algorithm theoretically runs in time proportional to $\mathcal{O}(n^{2.371552 k/3}) = \mathcal{O}(n^{0.7905173 k})$.
However, the bound for the matrix multiplication complexity exponent $\omega \leq 2.371552$ is known to be impractical in real-world computation, due to the astronomical constant coefficient hidden by the big O notation \cite{pan2017fast}.
The fastest practical fastest matrix multiplication algorithm \cite{pan1982trilinear} is such that $\omega = 2.7734$.
By contrast, the time complexity of the current best classical algorithm \cite{akagi2018exact} for the max-min dispersion problem is given by $\mathcal{O}(n^{\omega k/3}\log n) = \mathcal{O}(n^{0.7905173 k}\log n)$.

As shown in Section~\ref{subsec:gas_with_dicke}, the query complexity of the GAS with Dicke state is $\mathcal{O}\left(\sqrt{\binom{n}{k}/t}\right)$. Using the well known bound on the binomial coefficient $\binom{n}{k} < \qty(en/k)^k$, we find $\mathcal{O}\left(\sqrt{\binom{n}{k}/t}\right) \subset \mathcal{O}\left(\qty(en/kt)^{0.5k}\right)$, where $e$ is the base of the natural logarithm. This complexity is obviously smaller than the time complexities of the classical algorithms, which suggests the possibility of quantum speedup by future implementation of a scalable FTQC.

\section{Performance Analysis}
\label{sec:performance}

In this section, we evaluate the query complexity of the proposed method for both max-sum and max-min dispersion problems.
The results of the original GAS with Hadamard transform and the classical exhaustive search method are also presented as references.\footnote{Note that GAS can be regarded as a quantum exhaustive search algorithm.}
We consider the same two performance metrics adopted in \cite{botsinis2014fixedcomplexity, ishikawa2021quantum, yukiyoshi2022quantum}, namely, the query complexity in the quantum domain (QD), defined as the total number of oracle operators, \ie, $\sum_i L_i$; and the query complexity in the classical domain (CD), defined as the number of measurements of the quantum states.
Note that both query complexities in QD and CD are important metrics affecting the actual execution time of quantum algorithms.
In particular, the query complexity in QD is typically considered the primary metric, demonstrating quantum speedup over classical algorithms; while the query complexity in CD is also crucial, as the measurement process, including networking and circuit reconstruction with classical computation involved, can be a time-consuming task.

In our simulations, a total $10^4$ independent random distance matrices $(d_{i,j}) \in \mathbb{R}^{n \times n}$ are generated for each data point, with each distance $d_{i, j}$ drawn from a uniform integer distribution over the interval $[1, 20]$.
Using the generated data, the cumulative distribution function (CDF) and the median values for the query complexities are calculated and plotted.
In all results, we fix $n = 12$, and each result was presented for two cases with different Hamming weights: (a) $k = 6$, and (b) $k = 2$.
The availability of a sufficiently large number of register qubits $m$ was also assumed.

\subsection{Max-Sum Dispersion Problem}
\label{subsec:res:ma-sum}

\begin{figure}[H]
\centering
\subfigure[$(n, k) = (12, 6)$]{\includegraphics[width=\columnwidth]{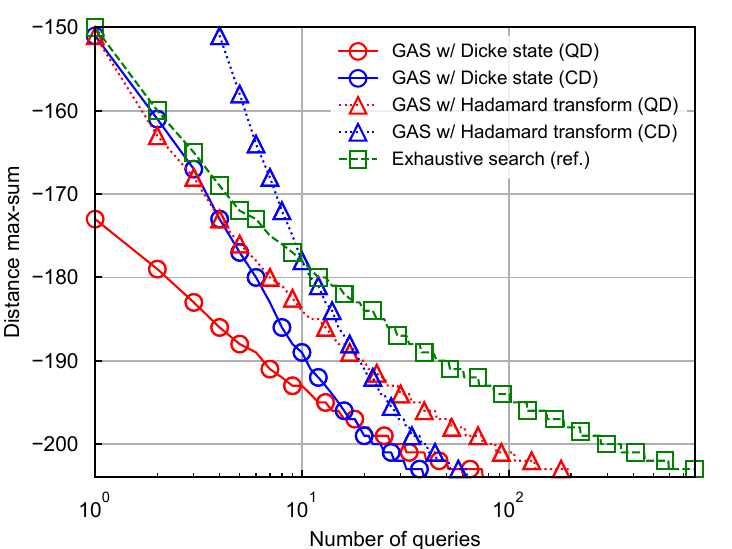}\label{fig:objective_max_sumk6}}
\subfigure[$(n, k) = (12, 2)$]{\includegraphics[width=\columnwidth]{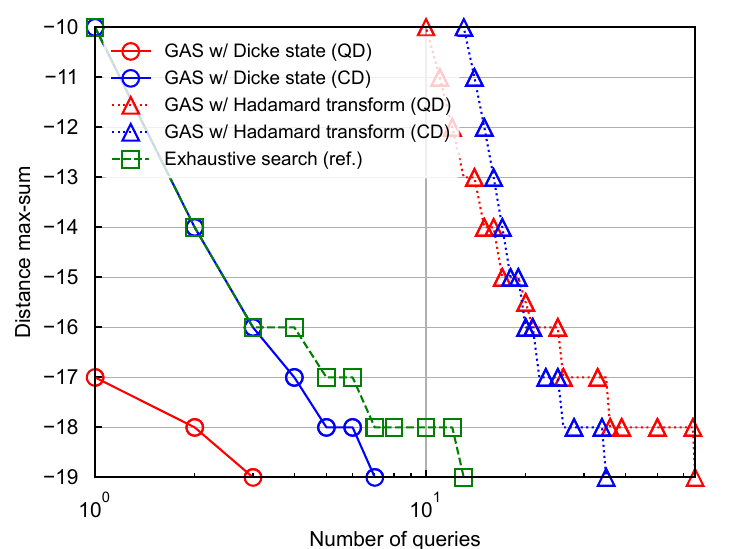}\label{fig:objective_max_sumk2}}
\caption{Relationship between the query complexity and the median of the objective function values in the max-sum dispersion problem.}
\label{fig:objective_max_sum}
\end{figure}

\begin{figure}[H]
\centering
\subfigure[$(n, k) = (12, 6)$]{\includegraphics[width=\columnwidth]{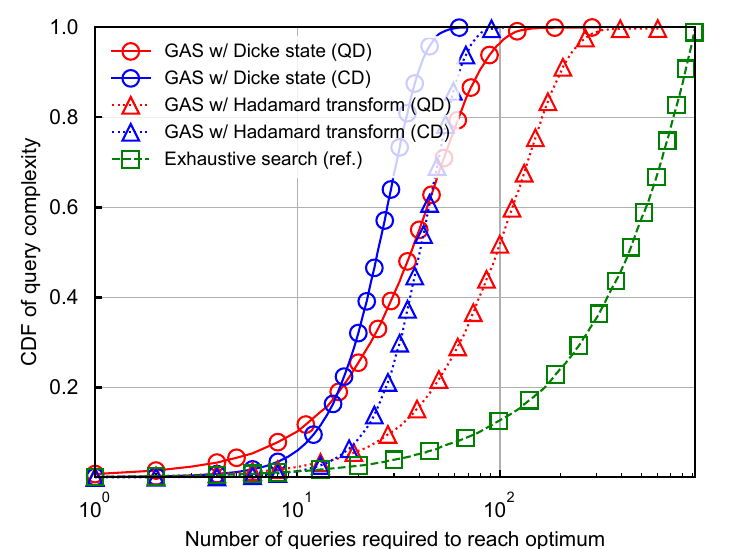}}
\subfigure[$(n, k) = (12, 2)$]{\includegraphics[width=\columnwidth]{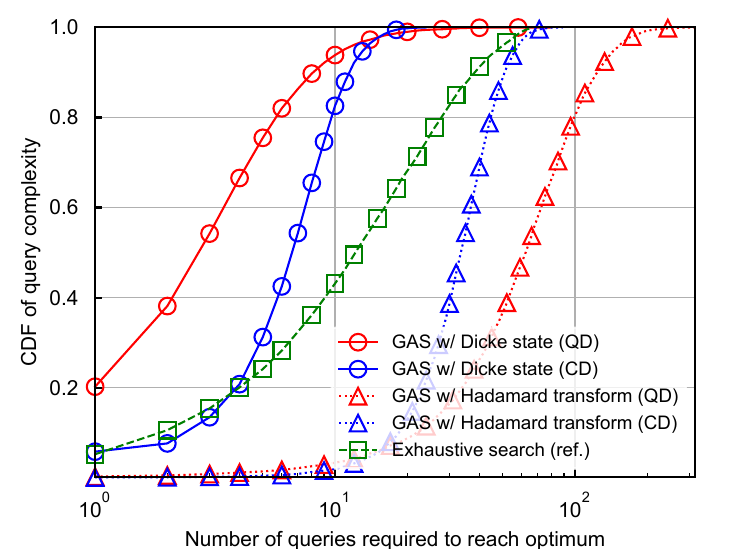}}
\caption{CDF of the query complexity required to reach the optimal solution in the max-sum dispersion problem.}
\label{fig:CDF_max_sum}
\end{figure}

We start by investigating the performance of the proposed method formulated in Section \ref{subsec:max-sum-DP} for the max-sum dispersion problem.
In all results, shown in Figs. \ref{fig:objective_max_sum} and \ref{fig:CDF_max_sum}, the penalty coefficient $\lambda_2$ in equation \eqref{eq:objfun_maxsum} was set to $\lambda_2 = 100$.
First, Fig. \ref{fig:objective_max_sum} shows the relationship between the query complexity and the median of the objective function values over $10^4$ trials.
The results indicate that the GAS with Dicke state exhibits the fastest convergence among the methods compared, both in QD and CD terms.

Notice that although the GAS with Hadamard transform achieved at least faster convergence than the classical exhaustive search in Fig.~\ref{fig:objective_max_sumk6}, with $k = 6$, much worse convergence than the classical exhaustive search was observed in Fig.~\ref{fig:objective_max_sumk2}, with $k = 2$, indicating the loss of quantum speedup in systems with small Hamming weights.
This can be clearly explained by comparing the theoretical query complexities.
Specifically, the query complexity of the GAS algorithm with Hadamard transform in QD can be larger than that of the classical exhaustive search when $k$ is relatively small, since $\sqrt{2^{12}/t} > \tbinom{12}{2}/t$ holds for $t \geq 2$.
\vspace{-2ex}

It was generally found, nevertheless, that the GAS algorithm with Dicke state always show faster convergence than the classical exhaustive search, regardless of configuration parameters, owing to its pure quadratic speedup.
Notice, in particular, that in Fig. \ref{fig:objective_max_sumk2}, the GAS algorithm with Hadamard transform exhibited extremely large objective function values in regions with a smaller number of queries, which can be attributed to the uniform superposition produced by the Hadamard transform, which includes the redundant search space, leading to higher objective function values that violate the penalty terms in \eqref{eq:objfun_maxsum}.

Next, given that GAS is a nondeterministic algorithm, its probabilistic performance was investigated in Fig. \ref{fig:CDF_max_sum}, which considered the CDF of the query complexity required to reach the optimal solution.
The results again confirm that the GAS algorithm with Dicke state succeeds in finding the optimal solution with the lowest query complexity in terms of both average and worst-case scenarios for both cases (a), with $k=6$, and (b), with $k=2$.
In contrast, the GAS algorithm with Hadamard transform required a larger query complexity, especially in the case (b), with $k=2$.

\subsection{Max-Min Dispersion Problem}
Finally, we assess the performance of the proposed method for the max-min dispersion problem formulated in Section \ref{sec:max-min-DP}.
In all results, the penalty coefficient $\lambda_2$ in equation \eqref{eq:objfun_maxmin} was set to $\lambda_2 = 1$, and the constant $\delta$ for the distance compression technique proposed in Section \ref{sec:max-min-DP} was set to $\delta = 10^{-5}$.
The results, shown in Fig. \ref{fig:objective_max_min}, elucidate the relationship between the query complexity and the median of the objective function values over $10^4$ trials.
Distance compression was applied in Figs. \ref{fig:objective_max_mink6} and \ref{fig:objective_max_mink2}, and not applied in Figs. \ref{fig:objective_max_mink6nocomp} and \ref{fig:objective_max_mink2nocomp}, in order to evaluate the effect of the technique.
As can be seen from Figs. \ref{fig:objective_max_mink6} and \ref{fig:objective_max_mink2}, the GAS with Dicke state achieves the fastest convergence in both QD and CD, consistent with the results of the max-sum dispersion problem given in Section \ref{subsec:res:ma-sum}.
Comparing in Fig. \ref{fig:objective_max_mink6} with \ref{fig:objective_max_mink6nocomp} and Fig. \ref{fig:objective_max_mink2} with \ref{fig:objective_max_mink2nocomp}, it is evident that the ranges of the objective function values are compressed, owing to the application of distance compression, which helps reducing the number of register qubits $m$ required to express objective function values, thus enhancing the feasibility of GAS using the proposed formulation.

\begin{figure}[H]
\centering
\subfigure[$(n, k) = (12, 6)$]{\includegraphics[width=\columnwidth]{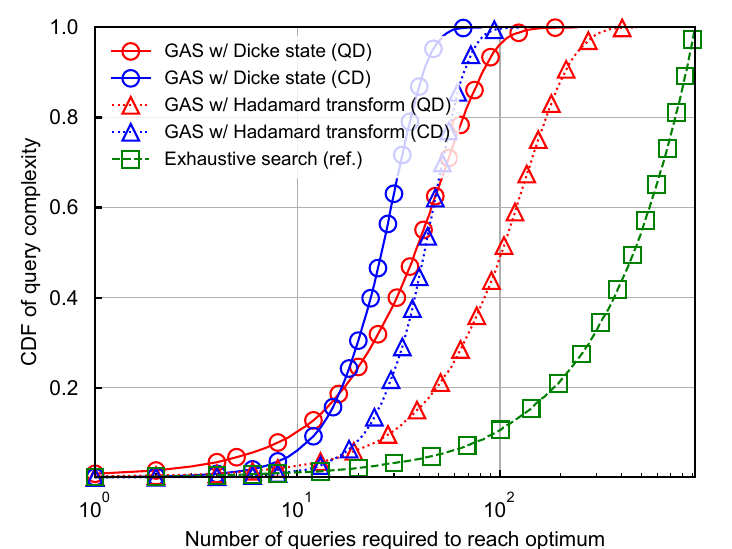}}
\subfigure[$(n, k) = (12, 2)$]{\includegraphics[width=\columnwidth]{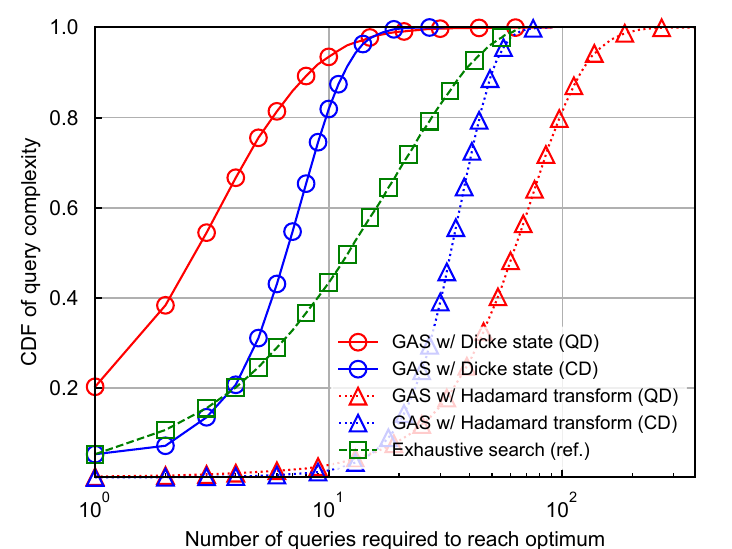}}
\caption{CDF of the query complexity required to reach the optimal solution in the max-min dispersion problem.}
\label{fig:CDF_max_min}
\end{figure}

Finally, Fig.~\ref{fig:CDF_max_min} compares the CDF of the query complexity required to reach the optimal solution with each of the schemes considered, showing the CDFs of the max-sum and max-min problem are almost identical.
\vspace{-3ex}

\begin{minipage}{1.0\textwidth}
\begin{figure}[H]
\centering
\subfigure[$(n, k) = (12, 6)$ with the distance compression technique applied.]{\includegraphics[width=0.45\columnwidth]{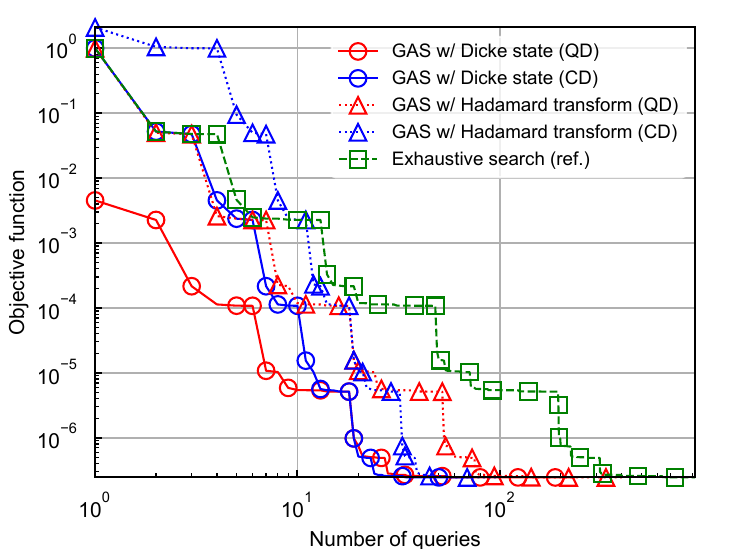}\label{fig:objective_max_mink6}}
\subfigure[$(n, k) = (12, 2)$ with the distance compression technique applied.]{\includegraphics[width=0.45\columnwidth]{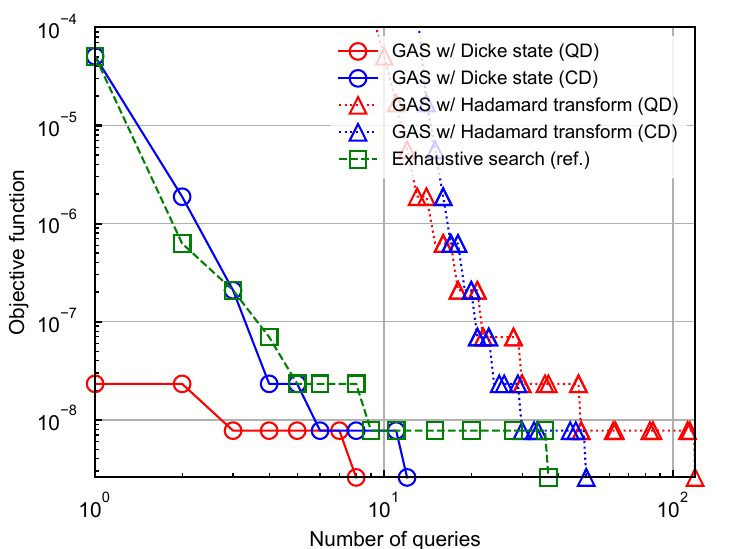}\label{fig:objective_max_mink2}}
\subfigure[$(n, k) = (12, 6)$ without the distance compression technique.]{\includegraphics[width=0.45\columnwidth]{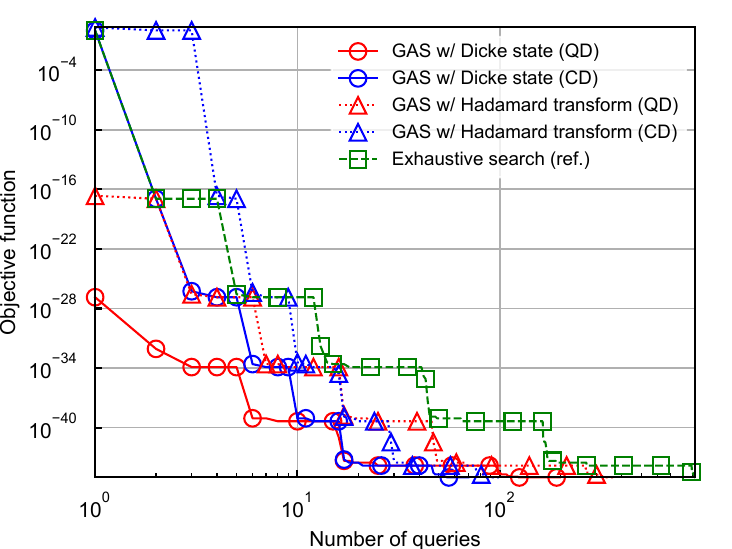}\label{fig:objective_max_mink6nocomp}}
\subfigure[$(n, k) = (12, 2)$ without the distance compression technique.]{\includegraphics[width=0.45\columnwidth]{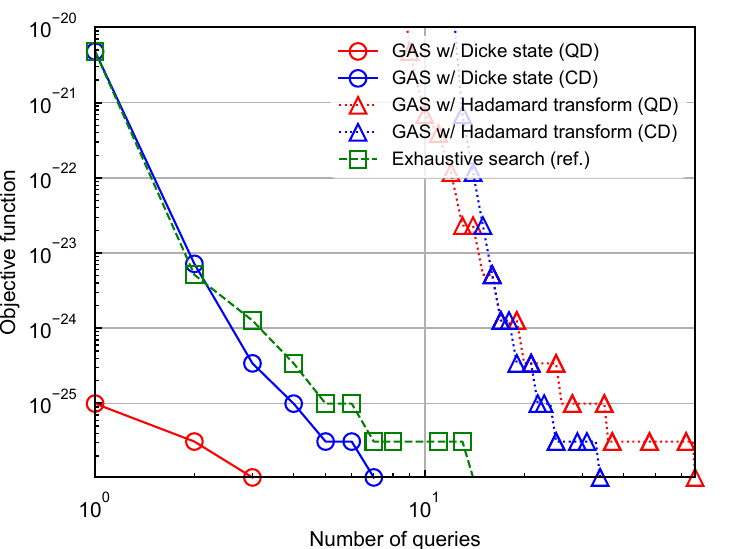}\label{fig:objective_max_mink2nocomp}}
\caption{Relationship between the query complexity and the median of the objective function values in the max-min dispersion problem.}
\label{fig:objective_max_min}
\end{figure}
\end{minipage}

\newpage

This similarity suggests that we can accurately estimate the query complexity performance of GAS based on the theoretical order of the query complexity, regardless of the problems to which it is applied.

\section{Conclusions}
\label{sec:conc}

We considered two classic dispersion problems, namely, the max-sum and max-min problems, and by extension the associated codebook design problems with vast application in wireless communications and coding theory.
For such problems, we formulated corresponding GAS algorithms incorporate Dicke states, enabling their solution via ML search with full quadratic quantum speedup.
It was shown that the search of an optimal solution to the dispersion problem over Dicke states significantly reduces the search space leading to a simplification of the quantum circuit via the elimination of penalty terms.
In addition, a mechanism to replace distance coefficients with corresponding relative ranks that preserve mutual amplitude relatioship was introduced, with enables a further reduction on the number of qubits required to implement the method.
An analysis was provided which demonstrated the potential reduction in query complexity compared to the conventional GAS using Hadamard transform, which were then confirmed numerically.

\newpage

\section*{Acknowledgment}
The authors would like to thank Prof. Keisuke Fujii and Prof. Kosuke Mitarai from Osaka University, Japan, for providing the potential idea of applying Dicke states to GAS.


\end{document}